\pdfoutput=1

\documentclass[12pt,reqno,preprint]{article}
\usepackage{jheppub}
\usepackage{amsmath,amssymb,amsfonts}
\usepackage{booktabs}
\usepackage{ifsym}
\usepackage{MnSymbol}
\usepackage[usenames,dvipsnames]{xcolor}
\usepackage{hyperref}
\hypersetup{
    colorlinks,
    urlcolor=Maroon,
    linkcolor=Maroon,
    citecolor=Bittersweet
}
\def \be {\begin{equation}}
\def \ee {\end{equation}}
\def \ba {\begin{eqnarray}}
\def \ea {\end{eqnarray}}
\def\nl{\nonumber\\}
\newcommand{\ui}[1]{\textrm{#1}}
\def\SL2C{\text{vol}\,\mathrm{SL}(2,\mathbb{C})}
\def\f{f}

\begin{document}
\title{One-Loop Corrections From Higher Dimensional Tree Amplitudes}

\author{Freddy Cachazo${}^{a}$, Song He${}^{b}$ and Ellis Ye Yuan${}^{c}$}
\affiliation[a]{Perimeter Institute for Theoretical Physics, Waterloo, ON N2L 2Y5, Canada}
\affiliation[b]{Institute of Theoretical Physics, Chinese Academy of Sciences, Beijing 100190, P.R.China}
\affiliation[c]{School of Natural Sciences, Institute for Advanced Study, Princeton, NJ 08540, USA}
\emailAdd{fcachazo@pitp.ca, songhe@itp.ac.cn, yyuan@ias.edu}

\abstract{
We show how one-loop corrections to scattering amplitudes of scalars and gauge bosons can be obtained from tree amplitudes in one higher dimension. Starting with a complete tree-level scattering amplitude of $n+2$ particles in five dimensions, one assumes that two of them cannot be ``detected" and therefore an integration over their LIPS is carried out. The resulting object, function of the remaining $n$ particles, is taken to be four-dimensional by restricting the corresponding momenta. We perform this procedure in the context of the tree-level CHY formulation of amplitudes. The scattering equations obtained in the procedure coincide with those derived by Geyer et al from ambitwistor constructions and recently studied by two of the authors for bi-adjoint scalars. They have two sectors of solutions: regular and singular. We prove that the contribution from regular solutions generically gives rise to unphysical poles. However, using a BCFW argument we prove that the unphysical contributions are always homogeneous functions of the loop momentum and can be discarded. We also show that the contribution from singular solutions turns out to be homogeneous as well.
}
\maketitle

\section{Introduction}

The leading order in a perturbative computation of scattering amplitudes in quantum field theory is given by summing over tree-like Feynman diagrams. Quantum mechanical corrections to these are given by diagrams with loops. Although superficially different, Feynman pointed out that loops are not necessarily new objects and that by using contour deformations and distributional arguments, one-loop diagrams can be related to trees in his famous Feynman tree theorem \cite{Feynman:1963ax} (also see \cite{Catani:2008xa,CaronHuot:2010zt} for some more recent developments).

In recent years there has been renewed interest in the reformulation of scattering amplitudes using only on-shell physics \cite{ArkaniHamed:2012nw}. This led to the on-shell diagram expansion of planar ${\cal N}=4$ super Yang--Mills (SYM) amplitudes at all-loop orders \cite{ArkaniHamed:2010kv,ArkaniHamed:2012nw} (see ~\cite{Boels:2010nw} for discussions on more general theories). The key ingredient in all these developments has been the construction of loop amplitudes using forward limits of lower loop amplitudes with additional on-shell particles.

In this work we show that starting with a tree-level scattering amplitude of $n+2$ massless particles in five dimensions it is possible to obtain, in a natural way, a one-loop correction to the scattering of $n$ particles in four dimensions (the discussion is restricted to four dimensions for concreteness). The procedure assumes that two of the particles cannot be ``detected'' and thus must be integrated out\footnote{The idea of ``hiding'' a pair of particles to form a loop was previously utilized in, e.g., all-loop recursion relations~\cite{ArkaniHamed:2010kv} and the amplituhedron \cite{Arkani-Hamed:2013jha}}. This is done in five dimensions by integrating over the corresponding Lorentz-invariant phase space of two massless particles (${\rm LIPS}_2$). The resulting object, function of the remaining $n$ particles, is still five dimensional. In order to have a four-dimensional amplitude, the momenta of the remaining $n$ particles are taken to lie in a four-dimensional subspace and a momentum conserving delta function is included to ensure that the object is physical. On the support of the new four-dimensional delta function the ${\rm LIPS}_2$ integration turns into a five-dimensional forward limit. This procedure is described in detail in Section \ref{sec2}.

In the rest of the paper, we apply this procedure to the CHY formulation of scattering amplitudes, both for bi-adjoint scalars and gauge bosons~\cite{Cachazo:2013gna,Cachazo:2013hca, Cachazo:2013iea, Cachazo:2014nsa,Cachazo:2014xea}.

In Section \ref{sec3} we show how higher-dimensional forward limits work in the CHY representation in general. The reason for using the CHY representation and not Feynman diagrams directly is that the former provides a way to regulate the forward limit which would otherwise be divergent. In the limit, the CHY representation decomposes an amplitude into three parts corresponding to contributions from different solution sectors of the tree-level scattering equations. The resulting equations at one loop exclude solutions in one of the sectors where the divergence of the forward limit is absorbed, rendering the expression finite. The remaining two sectors are named ``regular'' and ``singular'' following the terminology in \cite{He:2015yua}. The scattering equations at one loop were obtained in \cite{Geyer:2015bja,Geyer:2015jch} based on one-loop results \cite{Adamo:2013tsa, Casali:2014hfa} of ambitwistor string theory \cite{Mason:2013sva}. The same equations were also analyzed in terms of the forward limit of two massive particles in \cite{He:2015yua} based on the works \cite{Naculich:2014naa,Dolan:2013isa}.

In Section \ref{sec4} we discuss general features of the CHY representation of the higher dimensional forward limit. We prove that the contribution from the regular solutions always possesses an unphysical pole. The unphysical pole is extracted using the BCFW technique \cite{Britto:2004ap,Britto:2005fq} for any multiplicity. This procedure reveals that the contribution from the pole is a function that is homogenous in the loop momentum. We also show that the contribution from the singular solutions is a homogenous function as well. We illustrate these features using the bi-adjoint $\Phi^3$ theory.

These results lead us to define the physical integrand as an equivalence class of rational functions, where two of them are identified if they differ by a homogenous function of the loop momentum. This definition is inspired by the recent Q-cut construction of loop integrands~\cite{Baadsgaard:2015twa}, where it was argued that in schemes such as dimensional regulation such objects can be defined to integrate to zero.

In Section \ref{sec5} we extend this analysis to the scattering of gluons with a scalar running in the loop and then to a gluon in the loop. Our discussions are all done with bosons and therefore do not rely on supersymmetry. The analysis for gluons as external particles contains new features which the pure scalar case does not have. The most interesting one is that the contribution from the singular solutions is inherently ambiguous as shown directly by its definition using the forward limit.  Once again, regardless of the particular definition used, this contribution is homogenous in the loop momentum.

We conclude Section \ref{sec5} by noting that the CHY formulas for gluon one-loop amplitudes obtained from our higher-dimensional forward limit coincide with those proposed in the very recent work by Geyer et al \cite{Geyer:2015jch} using ambitwistor constructions. This is actually very convenient since their unitarity analysis of the regular solutions' contribution is useful in showing that these proposals give rise to the correct equivalence class of physical amplitudes. We end in Section \ref{sec6} with our conclusions.

\section{From 5D to 4D and From Trees to Loops}\label{sec2}

In this section we describe in detail the construction of $n$-particle one-loop amplitudes from $(n+2)$-particle tree amplitudes in one higher dimension. In order to keep the notation simple and the discussion concrete, we concentrate on the case when the target one-loop amplitude is four-dimensional while the parent tree-level one is five-dimensional. The discussion in this section does not assume any particular technique for the computation of the amplitudes. It is only starting in the next section that we use the CHY representation of tree amplitudes.

Let us start by considering a tree-level scattering amplitude of $n+2$ massless particles in five dimensions\footnote{In order to regulate divergent integrals the dimension can be turned into $(4-2\epsilon)+1$. Also, we believe that the restriction to massless particles in this section is not essential.},
\be\label{FiveD}
{\cal A}^{\rm tree,5D}_{n+2} = \delta^5(K_1+K_2+\cdots + K_{n} + K_{+} + K_{-})A_{n+2}^{\rm tree,5D}.
\ee
Here the $K$'s represent five-dimensional massless momentum vectors. For later convenience we label the last two particles by $+$ and $-$. At this point, apart from the requirement of masslessness, the theories under consideration are completely general.

Assuming that particles $+$ and $-$ escape ``detection'', it is natural to integrate them out. This is done by integrating over their corresponding Lorentz invariant phase space (LIPS) in order to yield a function that only depends on the first $n$ particles. The LIPS integration is done as follows
\be
H(1,2,\ldots ,n) = \int d^5K_{+}d^5K_{-}\delta(K_{+}^2)\delta(K_{-}^2){\cal A}^{\rm tree,5D}_{n+2}.
\ee
The delta functions are introduced to ensure that the two particles are on-shell. The function $H(1,2,\ldots ,n)$ only depends on the physical data of the remaining $n$ particles. For the moment we take particles $+$ and $-$ as scalars.

Preparing the ground for the reduction to four dimensions, it is useful to parameterize the momenta as
\begin{equation}\label{kin}
K_{\textsc a}^{\rm M} = (k_{\textsc a}^\mu;e_{\textsc a})\quad{\rm for}\quad \textsc{a}\in\{1,2,\ldots ,n,+,-\},
\end{equation}
where ${\rm M}\in\{0,1,2,3,4\}$ is a five-dimensional Lorentz index and $\mu\in\{0,1,2,3\}$ is a four-dimensional one; $e_{\textsc a}$ denotes the component in the fifth direction (which is spatial).

We follow the way to perform the dimensional reduction of amplitudes in momentum space explained in \cite{Cachazo:2013iaa}. Explicitly, an integral over every $e_a$ is performed, with a delta function setting $e_a=0$ for $a\in \{1,2,\ldots ,n-1\}$, but not for $a=n$. The reason is that momentum conservation in the fifth direction should impose the condition $e_n=0$ once the others are satisfied. Interestingly, the presence of particles $+$ and $-$ does not affect the argument.

Clearly, performing the restriction to four dimensions is still not enough to ensure that $H(1,2,\ldots ,n)$ is a physical amplitude since it has support on kinematical data that does not necessarily preserve four-dimensional momentum conservation. In order to produce a meaningful four-dimensional object it is necessary to include a four-dimensional momentum conserving delta function. It is this new object that becomes the one-loop correction to the four-dimensional scattering amplitude. Explicitly, our proposal for one-loop amplitudes reads
\be\label{prop}
{\cal A}^{\rm 1-loop,4D}_{n} =\delta^4(k_1+k_2+\cdots + k_{n})\int d^5K_{+}d^5K_{-}\delta(K_{+}^2)\delta(K_{-}^2) \widetilde{\cal A}^{\rm tree,5D}_{n+2},
\ee
where
\be
\widetilde{\cal A}^{\rm tree,5D}_{n+2} \equiv  \int \prod_{a=1}^nd e_a \prod_{b=1}^{n-1}\delta(e_b) {\cal A}^{\rm tree,5D}_{n+2}.
\ee

The starting point to make this connection is to rewrite the delta functions imposing the on-shell condition in \eqref{prop} as
\be\label{onshell}
\delta(K_{\pm}^2) = \delta(k_{\pm}^2-e_{\pm}^2) =\frac{1}{2e_{\pm}}\delta(|k_{\pm}|-e_{\pm})-\frac{1}{2e_{\pm}}\delta(|k_{\pm}|+e_{\pm}).
\ee
Here $|k_\pm|$ denotes the four-dimensional Minkowski norm. To be precise, the $\delta$ functions imposing the on-shell conditions such as the one in \eqref{onshell} are meant to be treated as poles in a contour integral. On the support of the four-dimensional momentum conservation, the five-dimensional $\delta$ function in \eqref{FiveD} becomes
\be\label{fdd}
\delta^5(K_1+K_2+\cdots + K_n + K_+ + K_-) = \delta^4(k_{+}+k_{-})\delta(e_{+}+e_{-}+e_n).
\ee
These delta functions imply that $|k_{+}|=|k_{-}|$ and therefore $\delta(K_{+}^2)\delta(K_{-}^2)$ gives rise to two different kind of supports. One of them gives $e_{+}=e_{-}$ while the other gives $e_{+}=-e_{-}$. Only the latter is compatible with the reduction to four dimensions which must get $e_n=0$ as a consequence of \eqref{fdd}.

Using the corresponding supports one finds
\be\label{strippedamp}
{\cal A}^{\rm 1-loop,4D}_{n} =\delta^4(k_1+k_2+\cdots + k_{n})\int \frac{d^4k_{+}d^4k_{-}}{|k_{+}||k_{-}|}\delta^4(k_{+}+k_{-})A^{\rm tree,5D}_{n+2}(\{(k_a;0)\},\{(k_{\pm};|k_{\pm}|)\}).
\ee
In order to write this formula in a more recognizable form, it is convenient to carry out the integration over $k_{-}$ and denote $k_{+}=\ell$,
\be\label{forwardlimit}
{\cal A}^{\rm 1-loop,4D}_{n} =\delta^4\big(\sum_{a=1}^{n}k_a\big)\int \frac{d^4\ell}{\ell^2}A^{\rm tree,5D}_{n+2}(\{(k_a;0)\},\{(\pm\ell;\pm|\ell|)\})
=:\delta^4\big(\sum_{a=1}^{n}k_a\big)\,A_n^{\rm 1-loop,4D}.
\ee
It is worth stressing the fact that the standard $1/\ell^2$ loop propagator here has a composite origin as $1/|k_{+}||k_{-}|$.

Note that even though $A_{n+2}^{\rm tree,5D}$ seems to depend on $|\ell|$, the fact that all other vectors are restricted to be four-dimensional implies that the only dependence on $\ell$ is in the form $k_a\cdot\ell$.

At first sight this result, \eqref{forwardlimit}, appears to be incorrect, since the only  standard loop Feynman propagator appearing is $1/\ell^2$. However, a simple but remarkable observation was made in \cite{Geyer:2015bja}: any one-loop Feynman diagram (and hence any one-loop amplitude) can be transformed into such a representation by applying partial-fraction relations, accompanied with proper shifts of the loop momentum.

Having found the final form of the proposal in the case of internal single scalar ( i.e. a scalar running in the loop),  we turn to the discussion of the more interesting case of internal particles with possibly flavor, color and polarization degrees of freedom.

Consider a theory where all particles are in the adjoint representation of a $U(N)$ group. This could be a flavor or a color group. The original five-dimensional amplitude can be written using the standard color decomposition as a sum over $(n+1)!$ terms
\be\label{trace}
\sum_{\omega\in S_{n+2}/\mathbb{Z}_{n+2}} {\rm Tr}\left( T^{a_{\omega_1}}T^{a_{\omega_2}}\cdots T^{a_{\omega_{n}}}T^{a_{\omega_{-}}}T^{a_{\omega_{+}}}\right){\cal A}^{\rm tree,5D}_{n+2}[\omega_1,\omega_2,\ldots ,\omega_{n},\omega_{-},\omega_{+}],
\ee
where the labels inside the square brackets denote the dependence on the ordering of partial amplitudes.
The forward limit of the kinematics requires that when summing over the $U(N)$ degrees of freedom of the two internal particles they must be identified. More explicitly, we introduce a $\delta_{a_{+}a_{-}}$ and sum over both sets of indices $a_{+}$ and $a_{-}$ from $1$ to $N^2$. Given that the generators $T^a$ form a basis of the space of $N\times N$ matrices, the sum fuses the trace indices at the locations of the relevant generators. This gives rise to two kinds of terms. The first comes from permutations $\omega$ such that particles $+$ and $-$ are adjacent. In this case we have
\be
\sum_{a_{+},a_{-}=1}^{N^2}\delta_{a_{+}a_{-}}{\rm Tr}\left( T^{a_{-}}T^{a_{+}}T^{a_{\omega_1}}T^{a_{\omega_2}}\cdots T^{a_{\omega_{n}}}\right) = N {\rm Tr}\left(T^{a_{\omega_1}}T^{a_{\omega_2}}\cdots T^{a_{\omega_{n}}}\right).
\ee
The second case gives rise to double-trace terms which are clearly present in $U(N)$ theories at one-loop but since they are determined by the single-trace terms~\cite{Bern:1996je}, we will not discuss them any further and concentrate instead on single-trace terms.

Clearly several original color orderings give rise to the same trace factor after summing over particles $+$ and $-$. Collecting terms according to the trace factor one finds that (\ref{trace}) becomes
\be\label{partial}
\sum_{\omega\in S_{n}/{\mathbb{Z}_n}} {\rm Tr}\left( T^{a_{\omega_1}}T^{a_{\omega_2}}\cdots T^{a_{\omega_{n}}}\right)\sum_{j=0}^{n-1}{\cal A}^{\rm tree,5D}_{n+2}[\omega_{1+j},\omega_{2+j},\ldots ,\omega_{n+j},-,+] ~ + ~ ({\rm double~traces}).
\ee

Now it is possible to put all elements together and write a single-trace one-loop partial amplitude with canonical ordering as
\be
{\cal A}^{\rm 1-loop,4D}_{n}[1,2,\ldots ,n]= \delta^4(k_1+k_2+\cdots + k_{n})A^{\rm 1-loop,4D}_n[1,2,\ldots ,n],
\ee
with
\be\label{fin}
A^{\rm 1-loop,4D}_n[1,2,\ldots ,n] = \int\frac{d^4\ell}{\ell^2}\sum_{j=0}^{n-1}A_{n+2}^{\rm tree,5D}[1+j,2+j,\ldots,n+j,-,+].
\ee
Here the arguments entering $A_{n+2}^{\rm tree,5D}$ are restricted in the same way as in \eqref{forwardlimit}, but for simplicity we suppress them.

Similar results apply to amplitudes with more symmetry groups such as the bi-adjoint scalar, which will be studied in Section \ref{sec4}. Here instead, we discuss the polarization degrees of freedom of the internal particles.

Let the polarization vectors of the two internal particles be $E_+^\ui{M}$ and $E_-^\ui{M}$ where the index $\ui{M}\in \{0,1,\ldots ,4\}$ (we reserve the notation $\epsilon_a^\mu$ for polarization vectors of external particles in four dimensions so that $E_a^{\rm M}=(\epsilon_a^\mu;0)$). Gauge bosons in five dimensions have three degrees of freedom and we denote physical polarizations by $E_{{\rm T}:i}$ where ``T" stands for transverse and $i\in \{1,2,3\}$. In the procedure to get one-loop amplitudes we sum over the possible physical polarizations, $E_+^{\ui{M}}$ and $E_-^\ui{M}$, subject to the constraint $E_-^\ui{M} = (E_+^\ui{M})^*$. This is possible since $K_+^\ui{M}=-K_-^\ui{M}$. Starting with the five-dimensional tree amplitude and making explicit the fact that it is a multi-linear function of $E_+^\ui{M}$ and $E_-^\ui{M}$, one has
\be\label{pola}
\sum_{E_+=(E_-)^*} A_{n+2}^{\rm tree,5D} = \sum_{i=1}^3 E_{+, {\rm T}:i}^{\ui{M}}(E_{+, {\rm T}:i}^{\ui{N}})^*A_{n+2; \ui{M}\,\ui{N}}^{\rm tree,5D}\,.
\ee
The five-dimensional completeness relation for polarization vectors expresses the sum in (\ref{pola}) as the metric, $\eta^{\ui{M}\,\ui{N}}$, up to two terms that contain longitudinal polarizations. These last two terms give vanishing contributions when contracted with the physical tree amplitude, thus we can write
\be\label{completeness}
\sum_{i=1}^3 E_{+, {\rm T}:i}^{\ui{M}}(E_{+,{\rm T}:i}^{\ui{N}})^*A_{n+2; \ui{M}\,\ui{N}}^{\rm tree,5D} = \eta^{\ui{M}\,\ui{N}}A_{n+2; \ui{M}\,\ui{N}}^{\rm tree,5D}.
\ee
Using this result the formula for a one-loop amplitude, where the internal particle is a $U(N)$ colored gauge boson, is given by
\be\label{gab}
A^{\rm 1-loop}_n[1,2,\ldots ,n] = \int\frac{d^4\ell}{\ell^2}\sum_{j=0}^{n-1}\eta^{\ui{M}\,\ui{N}}A_{n+2; \ui{M}\, \ui{N}}^{\rm tree,5D}[1+j,2+j,\ldots,n+j,-,+].
\ee
Here again we have not exhibited the explicit kinematic arguments in $A_{n+2}^{\rm tree,5D}$ in order not to clutter the notation but they are the same as those in \eqref{forwardlimit}.

\section{CHY Formulation and Regularization}\label{sec3}

As they stand now, our proposals \eqref{forwardlimit}, \eqref{fin} and \eqref{gab} are formal expressions since the corresponding RHS may suffer from divergences in the forward limit. In this section we show that by using the CHY representation of tree amplitudes these divergences can be tamed in a natural way.

Let us start with a brief review of the CHY construction. From now on we use indices $\textsc{a},\textsc{b},\ldots$ to run over labels $\{1,2,\ldots,n,+,-\}$, and $a,b,\ldots$ excluding the labels $\{+,-\}$. The CHY representation for a tree amplitude is given by
\be\label{massiveformula}
A^{\rm tree,5D}_{n{+}2}=\oint\frac{d^{n{+}2} \sigma_\textsc{a}}{\SL2C}\;\frac{I^{\rm tree,5D}_{n{+}2}}{\prod_{\textsc{a}}'\underline{F_{\textsc{a}}}},
\ee
where each $\sigma_{\textsc{a}}$ specifies the location of a puncture on a Riemann sphere associated to particle $\textsc{a}$, and the CHY integrand 
$I^{\rm tree,5D}_{n{+}2}(\{K,E,\sigma\})$ is a rational function whose explicit expression depends on the theory under study. The underline $\underline{F_{\textsc{a}}}$ indicates the contour $|F_{\textsc{a}}|=\varepsilon$, and
\be
F_{\textsc{a}}:=\sum_{{\textsc{b}}\neq {\textsc{a}}}\frac{K_{\textsc{a}}\cdot K_{\textsc{b}}}{\sigma_{\textsc{a}}-\sigma_{\textsc{b}}}\,,\qquad{\rm for}\quad \textsc{a}=1,2,\ldots, n,+,-\,,
\ee
are the functions imposing the tree-level scattering equations $\{F_{\textsc{a}}=0\}$. The integrals localize to solutions of the scattering equations, which are all non-degenerate for generic kinematic data and in our current setup count $(n-1)!$ in total \cite{Cachazo:2013gna,Dolan:2014ega}.

We follow the analysis in \cite{He:2015yua} and approach the forward limit by taking $K_{+}^{\rm M}+K_{-}^{\rm M}=\tau Q^{\rm M}$ and $\tau\to0$ ($Q^{\rm M}$ remains finite in the limit). It was observed in \cite{He:2015yua} that the $(n-1)!$ solutions separate into three sectors according to the behavior of the punctures $\sigma_{\pm}$ in the limit $\tau\to 0$, as summarized below
\begin{center}
\begin{tabular}{@{}c|cc@{}}
\toprule
solution sector & number of solutions & behavior \\
\midrule
regular & ~$(n-1)!-2(n-2)!$~ & $|\sigma_{+}-\sigma_{-}|\sim1$ \\
singular I & $(n-2)!$ & $|\sigma_{+}-\sigma_{-}|\sim\tau$ \\
~singular II~ & $(n-2)!$ & ~$|\sigma_{+}-\sigma_{-}|\sim\tau^2$~ \\
\bottomrule
\end{tabular}
\end{center}

As the behavior indicates, ``singular'' refers to the fact that the punctures $+$ and $-$ pinch in the limit, and the two types of singular sectors differ by the rate of pinching. Given that the CHY integrand for physical amplitudes usually comes with factors of the form $(\sigma_{\textsc a}-\sigma_{\textsc b})$ in its denominator, the singular solutions (especially those of type II)  may give rise to divergences, which is consistent with Feynman diagrams.

For example, for the forward limit of two scalars the leading scaling behavior of the contributions to the amplitude from different solutions are shown in the following table
\begin{center}
\begin{tabular}{@{}c|ccc@{}}
\toprule
$\displaystyle\substack{\text{Amplitudes with two}\\ \text{extra scalar particles}}$ & regular & singular I & singular II \\
\midrule
bi-adjoint $\Phi^3$ & $\tau^0$ & $\tau^0$ & $\tau^1$ \\
\midrule
gluon: all-plus & $\tau^0$ & $\tau^2$ & $\tau^5$ \\
gluon: one-minus & $\tau^0$ & $\tau^0$ & $\tau^1$ \\
gluon: MHV, NMHV, \ldots, $\overline{\rm MHV}$ & $\tau^0$ & $\tau^0$ & $\tau^{-1}$ \\
\bottomrule
\end{tabular}
\end{center}

The CHY formulation provides a systematic regularization for forward-limit divergences simply by imposing the limit on the tree-level formula prior to the integration over the puncture locations. In other words, one computes the forward limit of the CHY integrand as well as the contour, and then perform the $\sigma$ integration.

With this prescription the $n+2$ functions that impose the tree-level scattering equations reduce to
\ba\label{1loopsca}
&&F_a\Longrightarrow\f_a:=\sum_{b=1, b\neq a}^n  \frac{k_a \cdot k_b}{\sigma_{a}-\sigma_{b}}+\frac{k_a\cdot \ell}{\sigma_a{-}\sigma_+}-\frac{k_a\cdot \ell}{\sigma_a{-}\sigma_-}\,,\quad a\in \{1,2, \ldots , n\};\nl
&&F_{\pm}\Longrightarrow\f_\pm:=~\pm\sum_{b=1}^n \frac{\ell \cdot k_b}{\sigma_\pm-\sigma_b}.
\ea
The equations $\{f_A=0\}$ are known as the scattering equations at one loop, which were first obtained in \cite{Geyer:2015bja}.
Here we used the fact that $K_a^\ui{M}=(k^\mu_a;0)$ lie in a four-dimensional space, and $K_{+}=-K_{-}$ while staying on-shell, so $K_{+}\cdot K_{-}=0$.

As explained in \cite{He:2015yua}, the zero loci of \eqref{1loopsca} excludes the solutions in the singular II sector, and thus the $\sigma$ integral is now localized only to the remaining $(n{-}1)!-(n{-}2)!$ solutions, consisting of the regular sector and the singular I sector (where $\sigma_+=\sigma_-$). It turns out that for many theories of physical interest the divergences in the full tree-level amplitude only come from the integral localized on the singular II sector, hence the above prescription automatically regularizes the forward limit.

Furthermore, the one-loop CHY integrand is identified with the forward limit of the tree-level CHY integrand.  Following the discussion in Section \ref{sec2}, we sum over possible color indices and (for gauge bosons) polarizations of the two particles $+$ and $-$:\be\label{oneloopCHYI}
I^{\rm 1-loop,4D}_n=\sum_{a_+=a_-} \sum_{E_+=E_-^*}I^{\rm tree,5D}_{n{+}2}\big((k_1;0),\ldots, (k_n;0), (\ell;|\ell|), (-\ell;-|\ell|)\big)\,.
\ee
Therefore we obtain
\ba\label{1loopform}
A^{\rm 1-loop,4D}_n &=& \int \frac{d^4 \ell}{\ell^2}~\oint \frac{d\,^n\sigma_ad\sigma_+ d\sigma_-}{\SL2C}\;\frac{1}{\underline{\f_+}\,\underline{\f_-}\,\prod'_a\underline{\f_a}}
\;I^{\rm 1-loop,4D}_n,
\ea
with the one-loop CHY integrand as obtained from \eqref{oneloopCHYI}. In the rest of the paper we drop the explicit label ``1-loop,4D'' to simplify the notation.

Given the existence of two different solution sectors, it is natural to decompose the amplitude into two parts
\begin{equation}\label{Adecomposition}
A_n=A_n^{\text{reg.}}+A_n^{\text{sing.}},
\end{equation}
corresponding to the contributions from the localized integral in the two remaining sectors of solutions respectively. For later convenience, we define $\xi:=\sigma_+-\sigma_-$ and the combination
\begin{equation}
\f_++\f_-=\xi\sum_{b=1}^n\frac{\ell\cdot k_b}{\sigma_{+,b}\sigma_{-,b}}=:\xi \f_0,
\end{equation}
where $\sigma_{a,b}:=\sigma_a-\sigma_b$. Then the original contours in \eqref{1loopform} are equivalent to $|\f_a|=\varepsilon$, $|\f_+|=\varepsilon$ and $|\xi \f_0|=\varepsilon$. With this it is obvious that the regular solutions come from $f_0=0$ while the singular solutions from $\xi=0$. Thus it is justified to write
\begin{equation}\label{Areggeneral}
A_n^{\text{reg.}}=\int\frac{d^4\ell}{\ell^2}\oint\frac{d^n\sigma_ad\sigma_+d\xi}{\SL2C}
\frac{1}{\underline{\f_0}\;\underline{\f_+}\;(\prod'_a\underline{\f_a})}\frac{I_n}{\xi}.
\end{equation}
For $A_n^{\text{sing.}}$ one imposes $\underline{\xi}$ instead of $\underline{f_0}$.

\section{Interpretation of the Formula at One Loop}\label{sec4}

In the previous section we saw that the CHY formulation offers a natural regularization to the forward limit and that the two separated sectors of solutions of the scattering equations at one loop lead to the decomposition \eqref{Adecomposition}, i.e., $A_n=A_n^{\text{reg.}}+A_n^{\text{sing.}}$. This decomposition is not something one would naturally expect for a generic one-loop amplitude, say in a standard Feynman diagram computation, and so we need to gain a better understanding of these two parts as well as how to interpret them.

As usual, one checks the validity of \eqref{1loopform} as a formula for a one-loop amplitude by studying its consistency with unitarity. In doing this, we observe that on a generic unitarity cut or factorization, the residue at the corresponding kinematic pole only receives contributions from the regular part, $A_n^{\text{reg.}}$. Thus it is justified to suspect that the singular part, $A_n^{\text{sing.}}$, should somehow be physically irrelevant.

In order to study the contribution from the singular solutions it is convenient to explicitly write the analog of \eqref{Areggeneral}
\begin{equation}
\begin{split}
A_n^{\text{sing.}}=&\int\frac{d^4\ell}{\ell^2}\oint\frac{d^n\sigma_ad\sigma_+d\xi}{\SL2C}\frac{1}{\underline{\xi}\;\underline{\f_+}\;(\prod'_a\underline{\f_a^{\rm tree}})}\frac{I_n}{\f_0},
\end{split}
\end{equation}
where $\f_a^{\rm tree}=f_a\big|_{\xi=0}=\sum_{b\neq a}\frac{k_a\cdot k_b}{\sigma_{a,b}}$ impose the tree-level scattering equations. Clearly, all the $\f$'s are now homogeneous in $\ell^\mu$. Therefore, as long as $I_n$ is also homogeneous in $\ell^\mu$, $A_n^{\text{sing.}}$ is a scaleless integral. The implications of this fact are important and we discuss them below after considering $A_n^{\text{reg.}}$.

Let us now focus on the regular part $A_n^{\text{reg.}}$. The first observation is the possible presence of unphysical poles in the loop integrand of $A_n^{\text{reg.}}$. Indeed, from explicit calculations at four points it is known that the integrand of $A_4^{\text{reg.}}$ generically contains a pole of the form
\begin{equation}\label{badpole4pt}
(\ell\cdot k_1\,l\cdot k_2+\ell\cdot k_3\,\ell\cdot k_4)(k_1\cdot k_2)^2+(\ell\cdot k_1\,\ell\cdot k_3+\ell\cdot k_2\,\ell\cdot k_4)(k_1\cdot k_3)^2+(\ell\cdot k_1\,\ell\cdot k_4+\ell\cdot k_2\,\ell\cdot k_3)(k_1\cdot k_4)^2.
\end{equation}
It is natural to expect that this phenomenon continues to hold at higher multiplicity.

In the rest of this section we turn to the study of such unphysical poles and their contributions. We explicitly identify the unphysical pole that is present for any number of particles and extract its contribution via a BCFW argument. We show that as long as the CHY integrand is homogeneous in $\ell$ on the singular solutions, the potentially harmful pole is confined to a scaleless integral and hence it is rendered inoffensive.

When all of the above conditions are met and when unitarity cuts and factorization channels are also verified, it is then natural to define the physical one-loop amplitude as the equivalence class of objects where two are identified if they differ by a scaleless integral. This definition is the same as the one proposed in the recent Q-cut construction \cite{Baadsgaard:2015twa}. With the understanding that a physical integrand is an equivalence class, we can loosely write the following decomposition of our one-loop formulas
\begin{equation}\label{Adecomposition2}
A_n=A_n^{\rm physical}+A_n^{\rm scaleless}.
\end{equation}
Here the two parts are not necessarily unique but the decomposition is still useful. Moreover, as we will see in the next section, $A_n^{\text{sing.}}$ itself is ambiguous in the case of gluon scattering; however, this will turn out to be irrelevant as long as $A_n^{\text{sing.}}$ is always scaleless.

\subsection{Origin of the Unphysical Pole}

Whenever a pole is present in the amplitude and cannot be eliminated, it necessarily means that the amplitude has to blow up when kinematic data probes this pole. When viewed in terms of the formula, generically it indicates that on such singular kinematics some of the punctures should pinch in at least one solution to the scattering equations. At tree level and in most cases at loop level this pinch is associated to a physical pole, as has been extensively studied in the existing literature \cite{Cachazo:2013gna,Dolan:2013isa,Geyer:2015jch}.

However, for regular solutions at one loop, it is possible that the pinch $\xi=\sigma_{+,-}=0$ can occur on kinematic poles that are not physical\footnote{In \cite{Geyer:2015jch} it was shown that $\sigma_{+,-}\to0$ in the regular solutions occur when $\ell^\mu\to\infty$. But this limit is not a consequence of probing any pole explicitly present in the final result as what we are discussing here.}.
To see this, note that the two rational functions $\f_+=\f_+(\sigma_+)$ and $\f_-=\f_-(\sigma_-)$ have exactly the same functional expression. Let us denote the numerator of $\f_{\pm}$ as $N_{\pm}$; it is easy to see that the polynomial $N_{\pm}$ is of degree $n-2$ in $\sigma_{\pm}$, which generates $n{-}2$ roots $r_i=r_i(\sigma_a)$ for a given set of $\{\sigma_a\}$.

For any solution of the scattering equations, the variables $\sigma_+$ and $\sigma_-$ take values from these roots. Obviously, $\sigma_+$ and $\sigma_-$ must be the same root in the case of singular solutions and be different roots in the regular ones. Hence the situation that a regular solution becomes singular is equivalent to the situation that $N_\pm$ possesses degenerate roots. This happens when the discriminant of $N_\pm$ with respect to $\sigma_{\pm}$ vanishes
\begin{equation}\label{discN}
\text{Disc}(N_\pm)=0.
\end{equation}

Note that $\text{Disc}(N_\pm)$ is in general a function of $\{\sigma_a\}$, thus it is not obvious that \eqref{discN} can be possible. But recall that when $\sigma_+=\sigma_-$ the solutions of $\{\sigma_a\}$ are independent of the value of $\sigma_{\pm}$ and are identical to those from $\{\f_a^{\rm tree}=0\}$. One can see that as long as any of these tree-level solutions causes \eqref{discN} such singular solutions are allowed in the regular sector. This indicates that the combination
\begin{equation}\label{Delta}
\Delta:=\prod_{\{\sigma_a\}\,\in\,\text{tree soln.}}\text{Disc}(N_\pm)
\end{equation}
is the potential unphysical pole we are after. Note that $\Delta$ is guaranteed to be a rational function of the kinematics because the product is over all tree-level solutions. At four points this is exactly the unphysical pole \eqref{badpole4pt} noticed earlier.

Since this discussion does not rely on the form of CHY integrand $I_n$,  in general the unphysical pole \eqref{Delta} is present in the loop integrand of $A_n^{\text{reg.}}$.

\subsection{Homogeneity of Terms Containing the Unphysical Pole}\label{homogeneityproof}

An encouraging property of the unphysical pole \eqref{Delta} is that it is homogeneous in the loop momentum $\ell^\mu$, because its only $\ell^\mu$-dependence comes from the discriminant of $N_\pm$, which is homogeneous. Hence there is a possibility that $A_n^{\text{reg.}}$ is equivalent to an expression where the loop integrand is explicitly local. To confirm this, one has to show that a term in $A_n^{\text{reg.}}$ that contains the full dependence on the $\Delta$ pole can be extracted in the form of a scaleless integral. We now set up a general proof that this is indeed the case.

To extract the term that contains the $\Delta$ pole, a common procedure is to BCFW deform the external momenta by introducing  a parameter $z$, and so
\begin{equation}
A_n^{\text{reg.}}=\int\frac{d^4\ell}{\ell^2}\oint_{|z|=\varepsilon}\frac{dz}{z}A_n^{\text{reg.}}(z).
\end{equation}
We choose a deformation such that $\Delta=\Delta(z)$ also depends on $z$. By deforming the contour, the dependence of $A_n^{\text{reg.}}$ on $\Delta$ is fully captured by the residue at $z^*$ such that $\Delta(z^*)=0$ \footnote{This is true regardless of a possible pole at infinity.}. Thus we are justified to restrict our attention to such a contribution
\begin{equation}
A_n^{\text{reg.},\Delta}:=-\int\frac{d^4\ell}{\ell^2}\oint_{|\Delta(z)|=\varepsilon}\frac{dz}{z}A_n^{\text{reg.}}(z).
\end{equation}
In other words, the quantity $A_n^{\text{reg.}}-A_n^{\text{reg.},\Delta}$ is free of the $\Delta$ pole.

Now recall that in the $\sigma$ and $\xi$ integration in $A_n^{\text{reg.}}$ the contour is defined partly by $|\f_+|=\varepsilon$. On the support of $\f_+=0$, the condition $\Delta=0$ is equivalent to saying that $\f_+$ possesses degenerate roots, which is further equivalent to $\partial_+\f_+:=\frac{\partial}{\partial\sigma_+}\f_+=0$. Hence we have
\begin{equation}\label{ops}
A_n^{\text{reg.},\Delta}=-\int\frac{d^4\ell}{\ell^2}\oint_{|\partial_+\f_+(z)|=\varepsilon}\frac{dz}{z}\frac{d^n\sigma_ad\sigma_+d\xi}{\SL2C}\frac{1}{\underline{\f_0}\;\underline{\f_+}\;\prod'_a\underline{\f_a}}\frac{I_n}{\xi}.
\end{equation}
Note that $\f_0$ can be written as
\begin{equation}\label{jim}
f_0=-\partial_+\f_++\xi\,h(\xi),
\end{equation}
where $h(\xi)$ is some function such that $h(0)$ is finite, since $\lim_{\xi\to0}(\f_0+\partial_+\f_+)=0$.

Using \eqref{jim} to write $1/\xi$ as $h(\xi)/(f_0+\partial_+\f_+)$ one finds that \eqref{ops} becomes
\begin{equation}
A_n^{\text{reg.},\Delta}=-\int\frac{d^4\ell}{\ell^2}
\oint_{|\partial_+\f_+(z)|=\epsilon}\frac{dzd^n\sigma_ad\sigma_+d\xi}{\SL2C}
\frac{1}{\underline{\f_0}\;\underline{\f_+}\;\prod'_a\underline{\f_a}}\frac{I_n\,h(\xi)}{z\,(\f_0+\partial_+\f_+)}.
\end{equation}
The contour $\underline{\f_0}$ implies that $\f_0+\partial_+\f_+$ can be replaced by $\partial_+\f_+$ thus exhibiting an explicit pole in $\partial_+\f_+$. This means that the contour on integration $|\partial_+\f_+(z)|=\epsilon$ can be conveniently represented by our underline notation $\underline{\partial_+\f_+(z)}$ of the poles in the integrand,
\begin{equation}
A_n^{\text{reg.},\Delta}=-\int\frac{d^4\ell}{\ell^2}\oint\frac{dzd^n\sigma_ad\sigma_+d\xi}{\SL2C}
\frac{1}{\underline{\f_0}\;\underline{\f_+}\;(\prod'_a\underline{\f_a})\;\underline{\partial_+\f_+}}\frac{I_n\,h(\xi)}{z}.
\end{equation}
Finally, we perform another contour manipulation replacing $\underline{\f_0}$ by $h(\xi)\underline{\xi}$. This leads to
\begin{equation}\label{AregDelta}
\begin{split}
A_n^{\text{reg.},\Delta}=&-\int\frac{d^4\ell}{\ell^2}\oint\frac{dzd^n\sigma_ad\sigma_+d\xi}{\SL2C}\frac{1}{\underline{\xi}\;\underline{\f_+}\;(\prod'_a\underline{\f_a})\;\underline{\partial_+\f_+}}\frac{I_n}{z}\\
=&-\int\frac{d^4\ell}{\ell^2}\oint\frac{dzd^n\sigma_ad\sigma_+d\xi}{\SL2C}
\frac{1}{\underline{\xi}\;\underline{\f_+}\;(\prod'_a\underline{\f_a^{\rm tree}})\;\underline{\partial_+\f_+}}\frac{I_n}{z}.
\end{split}
\end{equation}
In the last equality we used that on the support of $\xi=0$ the functions $f_a$ turn into $f_a^{\rm tree}$.

As in the study of the singular contributions, note that the functions in $\{\f_a^{\rm tree}\}$ are independent of $\ell^\mu$. In addition, $\{\f_+,\partial_+\f_+\}$ are both linear in $\ell^\mu$, thus for any amplitude under study it suffices to verify that $I_n$ is homogeneous in $\ell^\mu$, on the support of the equations $\xi=f_a^{\rm tree}=\f_+=\partial_+f_+=0$. Once this last statement is verified for a given theory then one can conclude that the unphysical pole is harmless.

\subsection{A Warm-Up Example: The Bi-Adjoint $\Phi^3$ Amplitudes}

Now we use the bi-adjoint $\Phi^3$ theory as an illustrative example.

Tree-level amplitudes of the bi-adjoint $\Phi^3$ theory can be decomposed into double-partial amplitudes, which depend on two cyclic orderings as dictated by the flavor structure, $\alpha,\beta$~\cite{Cachazo:2013iea}\footnote{A different approach to one-loop $\Phi^3$ amplitudes was presented in \cite{Baadsgaard:2015hia}}. The CHY integrand for an $n$-point double-partial amplitude in this theory is $I_{\Phi^3,n}^{\text{tree}}[\alpha|\beta]=C^{\rm tree}[\alpha]\,C^{\rm tree}[\beta]$, where $C^{\rm tree}$ is the tree-level Parke--Taylor factor
\begin{equation}
C^{\rm tree}[\alpha]=\frac{1}{\sigma_{\alpha(1),\alpha(2)}\,\sigma_{\alpha(2),\alpha(3)}\,\cdots\,\sigma_{\alpha(n),\alpha(1)}}.
\end{equation}
Since there is no explicit dependence on the kinematic variables the forward limit is straightforward. From \eqref{fin} (and as explained also in \cite{He:2015yua}), the sum over color indices in \eqref{oneloopCHYI}  gives the one-loop Parke--Taylor factors with a further cyclic sum:
\begin{equation}\label{oneloopPhi3I}
I_{\Phi^3,n}[\alpha|\beta]=C[\alpha]\,C[\beta],\quad
C[1,2,\ldots,n]:=\sum_{\rho\,\in\,\mathbb{Z}_n}\,C^{\rm tree}[\rho(1),\rho(2),\ldots,\rho(n),-,+]\,.
\end{equation}
Thus there is no $\ell^\mu$ dependence in $I_n^{\Phi^3}$ at all, confirming that $A_n^{\rm sing.}$ is scaleless.

In \cite{He:2015yua}, it was argued that the forward limit of the tree-level CHY formula gives
\begin{equation}
A_{\Phi^3,n}=A_{\Phi^3,n}^{\text{F.D.}}-\frac{A_{\Phi^3,n}^{\text{tree}}}{4}\,\int\frac{d^4\ell}{\ell^2}\sum_{a=1}^n\frac{1}{(\ell \cdot k_a)^2},
\end{equation}
where $A_{\Phi^3,n}^{\text{F.D.}}$ is the one-loop amplitude as calculated from Feynman diagrams, and $A_{\Phi^3,n}^{\text{tree}}$ is the corresponding tree-level amplitude. The term at the end of the above equation is explicitly homogeneous in $\ell^\mu$, and so this is in complete agreement with the decomposition \eqref{Adecomposition2} we obtained at the beginning of the section.

To see the relation with the other decomposition \eqref{Adecomposition}, note that $A_{\Phi^3,n}^{\text{reg.}}$ contains the unphysical pole $\Delta$ but it is contained in a scaleless integral, again because $I_{\Phi^3,n}$ is independent of $\ell^\mu$. Thus $A_{\Phi^3,n}^{\text{F.D.}}$ and $A_{\Phi^3,n}^{\text{reg.}}$ are equivalent since they only differ by scaleless integrals.

\section{Gluon Scattering}\label{sec5}

Here we apply the analysis in the previous section to gluon scattering, for both the case of a scalar loop and the case of a gluon loop. We first show that both the contribution from the singular solutions and that containing the unphysical pole from the regular solutions, are again collected into terms homogeneous in $\ell^\mu$. Relations to the formulas obtained in \cite{Geyer:2015jch} are commented in the end.

\subsection{Formula for Gluon Scattering with a Scalar Loop}

We first focus on the scattering of $n$ gluons with an adjoint scalar running in the loop. At tree level we need the scattering of $n$ gluons with two additional massless adjoint scalars, which we again label by $+$ and $-$. The CHY formula was obtained in~\cite{Cachazo:2014xea}, and the integrand for a partial amplitude (say with canonical ordering) is
\be
I_{n_{\texttt{g}}+2_{\texttt{s}}}^{\text{tree}}[1,2,\ldots,n,-,+]=C^{\rm tree}[1,2,\ldots, n,-,+]\,\frac{1}{\sigma_{+,-}}\,\text{Pf}'[\Psi_{n+2}^{\rm tree}]_{:\hat{+},\hat{-}}
\ee
where the matrix $\Psi_{n+2}$ is an $2(n+2)\times2(n+2)$ anti-symmetric matrix (whose indices are in the label set $\{1,2,\ldots,n,+,-:1,2,\ldots,n,+,-\}$), with the block structure
\begin{equation}
\Psi_{n+2}^{\rm tree}:=\left(\begin{matrix}\mathrm{A}_{n+2}&-\mathrm{C}_{n+2}^{\rm T}\\ \mathrm{C}_{n+2}&\mathrm{B}_{n+2}\end{matrix}\right),
\end{equation}
where the entries are
\begin{equation}\begin{split}
&(\mathrm{A}_{n+2})_{\textsc{a},\textsc{b}}:=\begin{cases}\frac{K_{\textsc a}\cdot K_{\textsc b}}{\sigma_{\textsc{a},\textsc{b}}}&\textsc{a}\neq \textsc{b}\\ 0&\textsc{a}=\textsc{b}\end{cases},\quad
(\mathrm{B}_{n+2}) _{\textsc{a},\textsc{b}}:=\begin{cases}\frac{E_{\textsc a}\cdot E_{\textsc b}}{\sigma_{\textsc{a},\textsc{b}}}&\textsc{a}\neq \textsc{b}\\ 0&\textsc{a}=\textsc{b}\end{cases},\quad\\
&(\mathrm{C}_{n+2}) _{\textsc{a},\textsc{b}}:=\begin{cases}\frac{E_{\textsc{a}}\cdot K_{\textsc b}}{\sigma_{\textsc{a},\textsc{b}}}&\textsc{a}\neq \textsc{b}\\ -\sum_{\textsc{c}\neq \textsc{a}}C_{\textsc{a},\textsc{c}}&\textsc{a}=\textsc{b}\end{cases}.
\end{split}\end{equation}
Then the reduced Pfaffian is defined by
\begin{equation}
\text{Pf}'[\Psi_{n+2}^{\rm tree}]_{:\hat{+},\hat{-}}:=\frac{(-1)^{\textsc{a}+\textsc{b}}}{\sigma_{\textsc{a},\textsc{b}}}\text{Pf}[\Psi_{n+2}^{\rm tree}]_{\hat{\textsc a},\hat{\textsc b}:\hat{+},\hat{-}},
\end{equation}
where the notation $[\Psi_{n+2}^{\rm tree}]$ denotes a submatrix of $\Psi_{n+2}^{\rm tree}$, and the hat specifies the row/column to be removed, in the first and in the second block respectively (as indicated by the colon in the middle). The result is independent of the choice of $\{\textsc{a},\textsc{b}\}$.

As we go to one loop, $C^{\rm tree}$ is again replaced by its corresponding loop-level $C[1,2,\ldots,n]$ due to the color structure. The matrix $\Psi_n$ at one loop directly descends from $\Psi_{n+2}^{\rm tree}$ following \eqref{oneloopCHYI}, i.e.,  by restricting $E_{\textsc a}$ to the four-dimensional $\epsilon_{\textsc a}$ and substituting $k_+\to \ell$ and $k_-\to-\ell$. Note that $K_+\cdot K_-\to0$ and so the entry $(\mathrm{A}_{n+2})_{+,-}=(\mathrm{A}_{n+2})_{-,+}=0$, and also the diagonal elements $(\mathrm{C}_{n+2})_{a,a}:=-\sum_{b\neq a}\frac{\epsilon_a\cdot k_b}{\sigma_{a,b}}-\frac{\epsilon_a\cdot \ell}{\sigma_{a,+}}+\frac{\epsilon_a\cdot \ell}{\sigma_{a,-}}$. As a result, for one-loop scattering we have
\begin{equation}\label{Iscalarloop}
I_{n_{\texttt{g}}}^{\text{scalar loop}}:=C[1,2,\ldots,n]\frac{1}{\sigma_{+,-}}\text{Pf}'[\Psi_n]_{:\hat{+},\hat{-}}.
\end{equation}

Having obtained the explicit CHY integrand, now we can straightforwardly prove that $A_{n_{\texttt{g}}}^{\text{scalar loop, sing.}}$ is scaleless. Note that the only place that can violate the homogeneity of $I_{n_{\texttt{g}}}^{\text{scalar loop}}$ in $\ell$ is in the entries $(\mathrm{C}_{n+2})_{a,a}$. But when $\xi=0$ its dependence on $\ell$ exactly cancel away, and so we are left with a homogeneous CHY integrand.

\subsection{Ambiguity from the Singular Solutions}

One should take the above result of the integrand \eqref{Iscalarloop} with a grain of salt, because it is ambiguous when evaluated upon the singular solutions: the definition of the reduced Pfaffian relies on the matrix $[\Psi_n]_{:\hat{+},\hat{-}}$ having corank $2$ on the support of the one-loop scattering equations, which turns out not to be true on the singular solutions. Thus for singular solutions different choices of the two rows/columns to be removed yield different loop integrands.

To understand this ambiguity in more detail, we apply a similarity transformation $[\Psi_n]_{:\hat{+},\hat{-}}\mapsto \mathrm{S} [\Psi_n]_{:\hat{+},\hat{-}}\mathrm{S}^{\rm T}$ by a matrix
\begin{equation}
\mathrm{S}=\left(\begin{matrix}
\mathbb{I}_n&0&0&0\\
0&1&1&0\\
0&\sigma_+&\sigma_-&0\\
0&0&0&\mathbb{I}_n
\end{matrix}\right),
\end{equation}
where $\mathbb{I}_n$ denotes the identity matrix of size $n\times n$. In other words, we transform the $\pm$ rows/columns from the first block into convenient linear combinations. Let us rename the two rows/columns after this transformation as $0$ and $0'$. They are both explicitly proportional to $\xi$, and we are allowed to pull two powers of this factor out of the reduced Pfaffian. Calling the resulting matrix $\tilde{\Psi}_n$, we have the identity
\begin{equation}
\frac{1}{\sigma_{+,-}}\text{Pf}'[\Psi_n]_{:\hat{+},\hat{-}}=\frac{\xi^2}{\sigma_{+,-}^2}\text{Pf}'\tilde{\Psi}_n=\text{Pf}'\tilde{\Psi}_n.
\end{equation}
In the above, the extra power of $\sigma_{+,-}$ in the denominator of the middle expression comes from the fact $\mathrm{det}(\mathrm{S})=\sigma_{+,-}$. Explicitly, this new matrix has the form
\begin{equation}\label{tildePsi}
\tilde\Psi\,=\,\left(\begin{array}{c|cc|c}
(\mathrm{A}_{n+2})_{a,b} & \frac{k_a\cdot\ell}{\sigma_{a,+}\sigma_{a,-}} & \frac{\sigma_a\,k_a\cdot\ell}{\sigma_{a,+}\sigma_{a,-}} & (-\mathrm{C}_{n+2})_{a,b} \\
\hline
-\frac{\ell\cdot k_b}{\sigma_{+,b}\sigma_{-,b}} & 0 & 0 & -\frac{\ell\cdot \epsilon_b}{\sigma_{+,b}\sigma_{-,b}} \\
-\frac{\sigma_b\,\ell\cdot k_b}{\sigma_{+,b}\sigma_{-,b}} & 0 & 0 & -\frac{\sigma_b\,\ell\cdot \epsilon_b}{\sigma_{+,b}\sigma_{-,b}} \\
\hline
(\mathrm{C}_{n+2})_{a,b} & \frac{\epsilon_a\cdot\ell}{\sigma_{a,+}\sigma_{a,-}} & \frac{\sigma_a\,\epsilon_a\cdot\ell}{\sigma_{a,+}\sigma_{a,-}} & (\mathrm{B}_{n+2})_{a,b}
\end{array}\right).
\end{equation}
Due to the tree-level origin of the matrix and the similarity transformation in obtaining $\tilde{\Psi}_n$, one would conclude that it has two null vectors
\begin{equation}\label{eigenvectors}
(\underbrace{1,1,\ldots,1}_{n},\xi,0:\underbrace{0,0,\ldots,0}_{n}),\quad
(\sigma_1,\sigma_2,\ldots,\sigma_n,0,\xi:\underbrace{0,0,\ldots,0}_{n}),
\end{equation}
which is true only if $\sum_{a=1}^n (A_{n+2})_{a,0}$
, which equals $f_0$, is zero. However, for generic kinematic data we know from Section \ref{sec3} that $f_0=0$ only holds on regular solutions. Hence we immediately conclude that the matrix $\tilde{\Psi}_n$ no longer has corank $2$ on the singular solutions and so the reduced Pfaffian is ambiguous inside $A_n^{\text{sing.},\Delta}$.

\subsection{Homogeneity of $A_{n_{\texttt{g}}}^{\text{reg.},\Delta}$}

In contrast to the singular solutions, the formula for the integrand is completely well-defined when integrated around the regular solutions, since on the support of the regular solutions it can be verified that the matrix $\tilde{\Psi}_n$ indeed has corank $2$ and so the CHY integrand \eqref{Iscalarloop} is unique. In this case, using the form of the eigenvectors \eqref{eigenvectors} the quantity $\text{Pf}'\tilde{\Psi}_n$ can be defined in several equivalent ways
\begin{equation}\label{defrPf}
\text{Pf}'\tilde{\Psi}_n:=\frac{(-1)^{a+b}}{\sigma_{a,b}}\text{Pf}[\tilde{\Psi}_n]_{\hat{a},\hat{b}:}
\equiv\frac{-(-1)^{a}}{\sigma_a\xi}\text{Pf}[\tilde{\Psi}_n]_{\hat{a},\hat{0}:}
\equiv\frac{(-1)^a}{\xi}\text{Pf}[\tilde{\Psi}_n]_{\hat{a},\hat{0'}:}
\equiv\frac{1}{\xi^2}\text{Pf}[\tilde{\Psi}_n]_{\hat{0},\hat{0'}:}.
\end{equation}

Generically $A_{n_{\texttt{g}}}^{\text{reg.}}$ also contains the unphysical pole $\Delta$ that we described in Section \ref{sec4}, and we need to prove that $A_{n_{\texttt{g}}}^{\text{reg.},\Delta}$ is scaleless. For this purpose we follow the general proof discussed in Section \ref{homogeneityproof}, and it suffices to show that the integrand \eqref{Iscalarloop} (thus the reduced Pfaffian therein) is homogeneous in $\ell^\mu$.

One might worry that $\text{Pf}'\tilde{\Psi}$ is again ambiguous, due to the presence of the contour $|\xi|=\varepsilon$ in the final expression we obtained in \eqref{AregDelta}. Luckily, this is not the case here. From the previous discussion we notice that the two linear redundancies of the matrix $\tilde{\Psi}_n$ are restored for kinematic configurations such that $\f_0=0$ as well. On the support of $\xi=0$, this is further equivalent to $\partial_+\f_+=-\f_0=0$. We have both $|\xi|=\varepsilon$ and $|\partial_+\f_+|=\varepsilon$ in \eqref{AregDelta}, and so $\text{Pf}'\tilde{\Psi}$ in this case is well-defined.

Due to the freedom in defining the reduced Pfaffian, it is most convenient to choose the first definition in \eqref{defrPf} so that the CHY integrand does not appear to diverge around the pole $\xi=0$. Then the homogeneity of the CHY integrand holds for the same reason as that in the situation of the singular solutions. We thus conclude that $A_{n_{\texttt{g}}}^{\text{scalar loop, reg.},\Delta}$ is also scaleless.

\subsection{A Gluon in the Loop}

The derivation of a formula for one-loop amplitudes in pure Yang--Mills, i.e., with a gluon loop, follows that of the scalar loop. Instead of using the submatrix $[\Psi^{\rm tree}_{n+2}]_{:\hat{+},\hat{-}}$ we start with the full $\Psi_{n+2}^{\rm tree}$ matrix, and in the forward limit we identify $E_+$ and $E_-$ and sum over all on-shell polarizations. The resulting integrand at one loop is thus
\begin{equation}\label{YMCHYI}
I_{\text{YM},n}[1,2,\ldots,n]=C[1,2,\ldots,n]\,\sum_{E_+=E_-^*}\,\text{Pf}'\Psi_{n},
\end{equation}
where the $2(n+2)\times2(n+2)$ matrix $\Psi_{n}$ is obtained from $\Psi_{n+2}^{\rm tree}$ in the same way as before. As expected, the reduced pfaffian is again ambiguous on the singular solutions, but well-defined on the regular solutions.

To confirm that this integrand is homogeneous in $\ell^\mu$ when evaluated on the contour in the final expression of \eqref{AregDelta}, all the arguments in the case of a scalar loop go through. Therefore, we conclude that $A_{\text{YM},n}^{\text{reg.},\Delta}$ is also a scaleless integral.

More explicitly, one can express $\text{Pf}'\Psi_n=\frac{1}{\xi}\text{Pf}'[\Psi_n]_{\hat{+},\hat{-}:}$ and plug in the completeness relation as done in \eqref{completeness} to obtain
\begin{equation}\label{gluonloop}
\begin{split}
\sum_{E_+=E_-^*}\text{Pf}'\Psi_{n}
=&-4\text{Pf}'\tilde{\Psi}_n+\sum_{a<b}\frac{(-1)^{a+b}\sigma_{a,b}k_a\cdot k_b}{\sigma_{+,a}\sigma_{-,a}\sigma_{+,b}\sigma_{-,b}}\text{Pf}[\Psi_n]_{\hat{a},\hat{b},\hat{+},\hat{-}:\hat{+},\hat{-}}\\
+&\sum_{a,b=1}^n\frac{(-1)^{a+b+n}\sigma_{a,b}k_a\cdot\epsilon_b}{\sigma_{+,a}\sigma_{-,a}\sigma_{+,b}\sigma_{-,b}}\text{Pf}[\Psi_{n}]_{\hat{a},\hat{+},\hat{-}:\hat{b},\hat{+},\hat{-}}+\sum_{a<b}\frac{(-1)^{a+b}\sigma_{a,b}\epsilon_a\cdot\epsilon_b}{\sigma_{+,a}\sigma_{-,a}\sigma_{+,b}\sigma_{-,b}}\text{Pf}[\Psi]_{\hat{+},\hat{-}:\hat{a},\hat{b},\hat{+},\hat{-}},
\end{split}
\end{equation}
where the matrix  $\tilde{\Psi}_n$ is the same one as obtained in \eqref{tildePsi} and its reduced Pfaffian defined in \eqref{defrPf}. When restricting to the all-plus helicity sector of the external gluons, the last three summations vanish and the formula \eqref{gluonloop} reduces to the one for the case of a scalar loop, up to an overall constant factor.

\subsection{Relation to Ambitwistor Formulas}

The formulas for gluon scattering (both with a scalar loop and with a gluon loop) that we obtained above from doing forward limits in higher dimensions are identical to the ones obtained by Geyer et al in \cite{Geyer:2015jch} by isolating contributions from different spin structures on the ambitwistor string worldsheet. More specifically, the matrix $\Psi_n$ here is the same as the matrix shown in their eq.~3.24, and so the formula with a gluon loop \eqref{YMCHYI} is equivalent to their eq.~3.20. Also, it is straightforward to see that the integrand in eq.~3.15a used in their paper for the scalar loop is the same as our \eqref{Iscalarloop}.

This is not completely surprising, because in \cite{Geyer:2015bja,Geyer:2015jch} a global residue theorem is applied to localize the worldsheet, which at one loop is a torus, to the point of the moduli space where it degenerates into a sphere with two punctures identified. 

In \cite{Geyer:2015jch} it was confirmed that for gluon scattering, the contribution from the regular solutions has the correct residue in every unitarity cut and factorization channel. Based on these verifications and our previous analysis of both $A_n^{\text{sing.}}$ and the unphysical pole in $A_n^{\text{reg.}}$, we thus conclude that these formulas are equivalent to physical amplitudes up to scaleless integrals.

\section{Conclusions}\label{sec6}

In this paper we showed how one-loop $n$-particle amplitudes can be obtained from tree-level $(n{+}2)$-particle amplitudes in one higher dimension. The procedure involves a dimensional reduction and an integration over the Lorentz-invariant phase space of two massless particles. Although we focused on one-loop amplitudes in four dimensions, the procedure clearly applies to general spacetime dimensions.

As a consequence of imposing the restriction that external particles satisfy momentum conservation, the internal particles are forced into a forward limit. Such limits are known to be plagued with divergences when applied to Feynman diagrams. However, we have shown that the CHY formulation of tree amplitudes provides a natural regularization and produces finite results. 

We argued that the loop integrand formulas thus obtained have to be interpreted as equivalent classes of expressions describing physical one-loop amplitudes modulo integrals that are scaleless. This definition is the same as the one given recently in the so-called Q-cut construction~\cite{Baadsgaard:2015twa} (see~\cite{Huang:2015cwh} for detailed discussions at one loop).

Modding out by scaleless integrals is necessary given the fact that the contribution from singular solutions can be ambiguous. We explicitly identified the source of the ambiguity in amplitudes with gluons and proved that the corresponding contribution is homogenous in the loop momentum. Moreover we showed that the contribution from regular solutions contains an unphysical pole. Hence simply discarding singular solutions is not enough to eliminate the need for the modding out procedure. Luckily, using a BCFW argument one can produce a formula that is free of the unphysical pole and in the same equivalence class as the original one.

The idea of one-loop amplitudes from forward limit of trees is not new \cite{Feynman:1963ax,Catani:2008xa,CaronHuot:2010zt, ArkaniHamed:2010kv,ArkaniHamed:2012nw}, but a main novelty here is to see how the forward limit comes from integrating out two hidden particles in one higher dimension; in this new procedure for one-loop amplitudes, the only physical propagator $\frac{1}{\ell^2}$ has a ``composite" origin from the dLIPS${}_2$ measure.

We used the procedure for a bi-adjoint scalar theory and pure Yang--Mills, but it is straightforward to consider other theories. For example, by considering gravitons in five dimensions, one expects to obtain a one-loop formula for gravity. Such a formula should agree with the one proposed in \cite{Geyer:2015jch}. We leave these questions to future work.

\acknowledgments
We thank K.~Ohmori for discussions and comments on the draft.
S.H. thanks B.~Feng for discussions and hospitality during his visit to Zhejiang University. E.Y.Y.~is supported by the U.S.~Department of Energy under grant DE-SC0009988, and by a Corning Glass Works Foundation Fellowship Fund at the Institute for Advanced Study. Research at Perimeter Institute is supported by the Government of Canada through Industry Canada and by the Province of Ontario through the Ministry of Research \& Innovation.

\bibliographystyle{JHEP}
\bibliography{HDFL}

\end{document}